\documentclass{svproc}
\usepackage[nosort]{cite}
\usepackage{amsfonts}
\usepackage{lastpage}
\usepackage{graphicx}
\usepackage{color}
\usepackage{inputenc}
\usepackage{comment}
\usepackage[T1]{fontenc}
\usepackage[inline]{enumitem}
\usepackage{float}
\usepackage{subcaption}
\usepackage{placeins}
\usepackage{afterpage}
\usepackage{etoolbox}
\patchcmd{\maketitle}{\@copyrightspace}{}{}{}
\usepackage{soul,color}
\usepackage{listings}
\usepackage[usenames,dvipsnames,svgnames,table]{xcolor}
\usepackage{todonotes}
\usepackage{hyperref}
\hypersetup{colorlinks=false}
\usepackage{array}
\newcolumntype{L}[1]{>{\raggedright\let\newline\\\arraybackslash\hspace{0pt}}m{#1}}
\newcolumntype{C}[1]{>{\centering\let\newline\\\arraybackslash\hspace{0pt}}m{#1}}
\newcolumntype{R}[1]{>{\raggedleft\let\newline\\\arraybackslash\hspace{0pt}}m{#1}}
\usepackage{listings}

\usepackage[group-separator={\,},  number-unit-product={\,}]{siunitx}
\usepackage{glossaries}
\usepackage{comment}
\makeglossaries
\newacronym{MITM}{MITM}{man-in-the-middle}
\newacronym{XTEA}{XTEA}{eXtended TEA}
\newacronym{RE}{RE}{reverse engineering}
\newacronym{BGA}{BGA}{ball grid array}
\newacronym{SWD}{SWD}{serial wire debug}
\newacronym{JTAG}{JTAG}{joint test action group}
\newacronym{PCB}{PCB}{printed circuit board}
\usepackage{afterpage}
\usepackage{ntheorem}
\usepackage{theorem} 
\theoremstyle{break}
\newtheorem{suggestion}{Suggestion}
\usepackage{url}

\makeatletter 
\g@addto@macro\UrlBreaks{\do\-}
\makeatother

\begin{document}
\mainmatter              
\title{Breaking Fitness Records without Moving: Reverse Engineering and Spoofing Fitbit}
\titlerunning{Breaking Fitness Records without Moving}  
\author{Hossein Fereidooni\inst{1} \and Jiska Classen\inst{2} \and Tom Spink \inst{3} \\ Paul Patras \inst{3} \and Markus Miettinen\inst{2} \\ Ahmad-Reza Sadeghi\inst{2} \and
Matthias Hollick\inst{2} \and Mauro Conti\inst{1} }
\authorrunning{H. Fereidooni et al.} 
\tocauthor{Hossein Fereidooni, Jiska Classen, Tom Spink, Paul Patras,
Markus Miettinen,  Ahmad-Reza Sadeghi, Matthias Hollick, and Mauro Conti}
\institute{University of Padua, Italy\\
\email{\{hossein,conti\}@math.unipd.it}\\ 
\and
Technische Universit\"at Darmstadt, Germany\\
\email{\{markus.miettinen,ahmad.sadeghi\}@trust.tu-darmstadt.de}\\ 
\email{\{jclassen,mhollick\}@seemoo.de}\\ 
\and
University of Edinburgh, United Kingdom \\
\email{\{tspink,ppatras\}@inf.ed.ac.uk} }
\maketitle              


\begin{abstract}
Tens of millions of wearable fitness trackers are shipped yearly to consumers who routinely collect information about their exercising patterns. Smartphones push this health-related data to vendors' cloud platforms, enabling users to analyze summary statistics on-line and adjust their habits. Third-parties including health insurance providers now offer discounts and financial rewards in exchange for such private information and evidence of healthy lifestyles. Given the associated monetary value, the authenticity and correctness of the activity data collected becomes imperative. In this paper, we provide an in-depth security analysis of the operation of fitness trackers commercialized by Fitbit, the wearables market leader. We reveal an intricate security through obscurity approach implemented by the user activity synchronization protocol running on the devices we analyze. Although non-trivial to interpret, we reverse engineer the message semantics, demonstrate how falsified user activity reports can be injected, and argue that based on our discoveries, such attacks can be performed at scale to obtain financial gains. We further document a hardware attack vector that enables circumvention of the end-to-end protocol encryption present in the latest Fitbit firmware, leading to the spoofing of valid encrypted fitness data. Finally, we give guidelines for avoiding similar vulnerabilities in future system designs.
\end{abstract}



\section{Introduction}

Market forecasts indicate 274 million wrist-based fitness trackers and smartwatches will be sold worldwide by 2020~\cite{forbes:2016}. Such devices already enable users and healthcare professionals to monitor individual activity and sleep habits, and underpin reward schemes that incentivize regular physical exercise. Fitbit maintains the lead in the wearables market, having shipped more units in 2016 than its biggest competitors Apple, Garmin, and Samsung combined~\cite{idc:2017}.

Fitness trackers collect extensive information which enables infering the users' health state and may reveal particularly sensitive personal circumstances. For instance, one individual recently discovered his wife was pregnant after examining her Fitbit data~\cite{mashable}. Police and attorneys start recognizing wearables as ``black boxes'' of the human body and use
statistics gathered by activity trackers as admissible evidence in court~\cite{wsj:2016,guardian:2014}.
These developments highlight the critical importance of both preserving data privacy throughout the collection process, and ensuring correctness and authenticity of the records stored. 
The emergence of third-party services offering rewards to users who share personal health information further strengthens the significance of protecting wearables data integrity. These include 
health insurance companies that provide discounts to customers who demonstrate physical activity through their fitness tracker logs \cite{vitalityhealth}, 
websites that financially compensate active users consenting to fitness monitoring \cite{achievemint}, and platforms where players bet on reaching activity goals to win money~\cite{stepbet}. Unfortunately, such on-line services also bring \emph{strong incentives for malicious users to manipulate tracking data, in order to fraudulently gain monetary benefits}.

Given the value fitness data has towards litigation and income, researchers have analyzed potential security and privacy vulnerabilities specific to activity trackers \cite{Rahman:2013,Britt,Eric:2016,Schellevis}.
Following a survey of 17 different fitness trackers available on the European market in Q1 2016~\cite{Fereidooni:2017}, recent investigations into the security of Fitbit devices (e.g.~\cite{Schellevis}), and the work we present herein, we found that in comparison to other
vendors, Fitbit employs the most effective security mechanisms in their products. 
Such competitive advantage, giving users the ability to share statistics with friends, and the company's overall market leadership make Fitbit one of the most attractive vendors to third parties running fitness-based financial reward programs. At the same time it motivates us to choose Fitbit trackers as the target of our security study, in the hope that understanding their underlying security architecture can be used to inform the security and privacy of future fitness tracker system designs.
Rahman \emph{et al.} have investigated the communication protocols used by early Fitbit wearables when synchronizing with web servers and possible attacks against this~\cite{Rahman:2013}. 
Cyr \emph{et al.}~\cite{Britt} studied the different layers of the Fitbit Flex ecosystem and argued correlation and \gls{MITM} attacks are feasible. Recent work documents firmware vulnerabilities found in Fitbit trackers~\cite{Eric:2016}, and the reverse engineering of cryptographic primitives and authentication protocols~\cite{Schellevis}.
However, as rapid innovation is the primary business objective, security considerations remain an afterthought rather than embedded into product design. Therefore, wider adoption of wearable technology is hindered by distrust~\cite{accenture,pwc}.

\textbf{Contributions}: We undertake an in-depth security analysis of the Fitbit Flex and Fitbit One fitness trackers and reveal serious security and privacy vulnerabilities present in these devices which, although difficult to uncover, are reproducible and \textbf{can be exploited at scale} once identified. Specifically, we reverse engineer the primitives governing the communication between trackers and cloud-based services, implement an open-source tool to extract sensitive personal information in human-readable format, and demonstrate that malicious users can inject fabricated activity records to obtain personal benefits. To circumvent end-to-end protocol encryption implemented in the latest firmware, we perform hardware-based \gls{RE} and document successful injection of falsified data that appears legitimate to the Fitbit cloud.  
The weaknesses we uncover, as well as the design guidelines we provide to ensure data integrity, authenticity and confidentiality, build foundations for more secure hardware and software development, including code and build management, automated testing, and software update mechanisms. 
Our insights provide valuable information to researchers and practitioners about the detailed way in which Fitbit operates their fitness tracking devices and associated services. These may help IoT manufacturers in general to improve their product design and business processes, towards developing rigorously secured devices and services.


\par{\textbf{Responsible Disclosure: }}
We have contacted Fitbit prior to submitting our work, and informed the company about the security vulnerabilities we discovered. We disclosed these vulnerabilities to allow sufficient time for them to fix the identified problems before the publication of our findings.
At the time of writing, we are aware that the vendor is in the process of evaluating the disclosed vulnerabilities and formulating an effective response to them.

\begin{figure}
\vspace{-2.2em}
\centering
  \includegraphics[width=.8\linewidth]{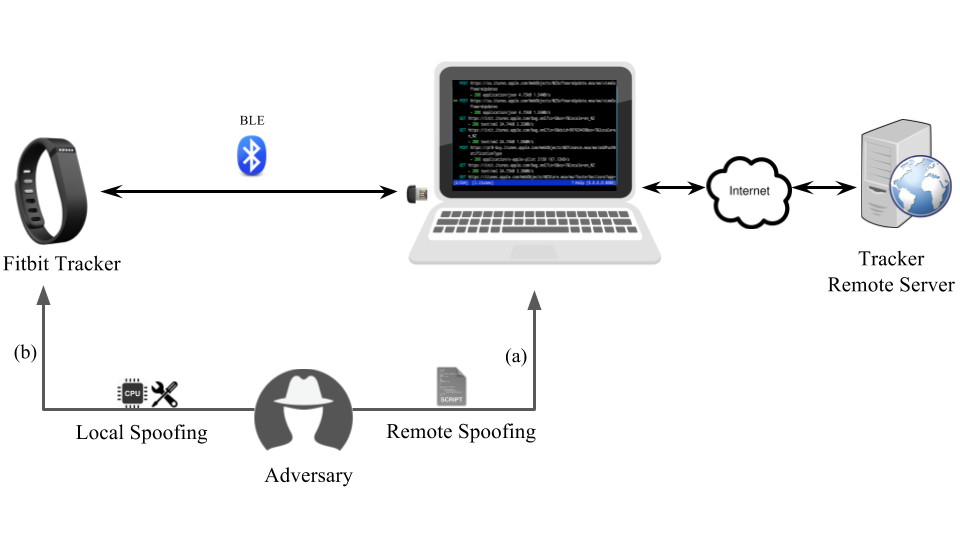}
  \vspace{-1em}
  \caption{Adversary model considered for (a) devices not implementing encryption and (b) trackers using encryption.}
  \label{fig:adversary}
\vspace{-1.5em}
\end{figure}

\section{Adversary Model} \label{sec:model}
To retrieve the statistics that trackers collect, users predominantly rely on smartphone or tablet applications that extract activity records stored by the devices, and push these onto cloud servers. We consider the adversarial settings depicted in \autoref{fig:adversary}, in which users are potentially dishonest, whilst the server is provably trustworthy. We assume an active adversary model in which the wristband user is the primary adversary, who has both the means and motive to compromise the system. Specifically, the attacker (a)~views and seeks to manipulate the data uploaded to the server without direct physical control over the device, or (b)~inspects and alters the data stored in memory prior to synchronization, having full hardware control of the device.
The adversary's motivation is rooted in the potential to obtain financial gains by injecting fabricated fitness data to the remote server. Smartphone and cloud platform security issues are outside the of scope of this paper, therefore not considered in our analysis.

\subsection{Target Fitbit Devices}
The adversary's target devices are the \emph{Fitbit Flex}  and \emph{Fitbit One } wrist-based fitness trackers, which record user step counts, distance traveled, calories burned, floors climbed (Fitbit One), active minutes, and sleep duration.
These particular trackers have been on the market for a number of years, they are affordable and their security and privacy has been scrutinized by other researchers. Thus, both consumers and the vendor would expect they are not subject to vulnerabilities. 

We subsequently found that other Fitbit models (e.g. Zip and Charge) implement the same communication protocol, therefore may be subject to the same vulnerabilities we identify in this work.

\subsection{End-to-End Communication Paradigms}

Following initial pairing, we discover Fitbit trackers are shipped with one of two different firmwares; namely, the latest version (Flex 7.81) which by default encrypts activity records prior to synchronization using the XTEA algorithm and a pre-installed encryption key; and, respectively, an earlier firmware version (Flex 7.64) that by default operates in plaintext mode, but is able to activate message encryption after being instructed to do so by the Fitbit server.
If enabled, \emph{encryption is end-to-end} between the tracker and the server, whilst the smartphone app is unaware of the actual contents pushed from tracker to the server. The app merely embeds encrypted records retrieved from the tracker into JSON messages, forwards them to the Fitbit servers, and relays responses back to the tracker. The same functionality can be achieved through software running on a computer equipped with a USB Bluetooth LE dongle, including the open-source Galileo tool, which does not require user authentication~\cite{Galileo}. 

Even though only the tracker and the server know the encryption key, upon synchronization the smartphone app also receives statistic summaries from the server in human readable format over an HTTPS connection. As such, and following authentication, the app and authorized third parties can connect to a user account via the Fitbit API and retrieve activity digests---without physical access to the tracker. We also note that, despite newer firmware enforcing end-to-end encryption, the Fitbit server continues to accept and respond to unencrypted activity records from trackers that only optionally employ encryption, thereby enabling an attacker to successfully modify the plaintext activity records sent to the server. 


\section{Protocol Reverse Engineering}
\label{sec:protocol-format}

In this section, we reverse engineer the communication protocol used by the Fitbit trackers studied, uncovering an intricate security through obscurity approach in its implementation. Once we understand the message semantics, we show that detailed personal information can be extracted and
fake activity reports can be created and remotely injected, using an approach that scales, as documented in \autoref{sec:protocol}.

\begin{figure*}[t]
\centering
  \includegraphics[width=.9\linewidth]{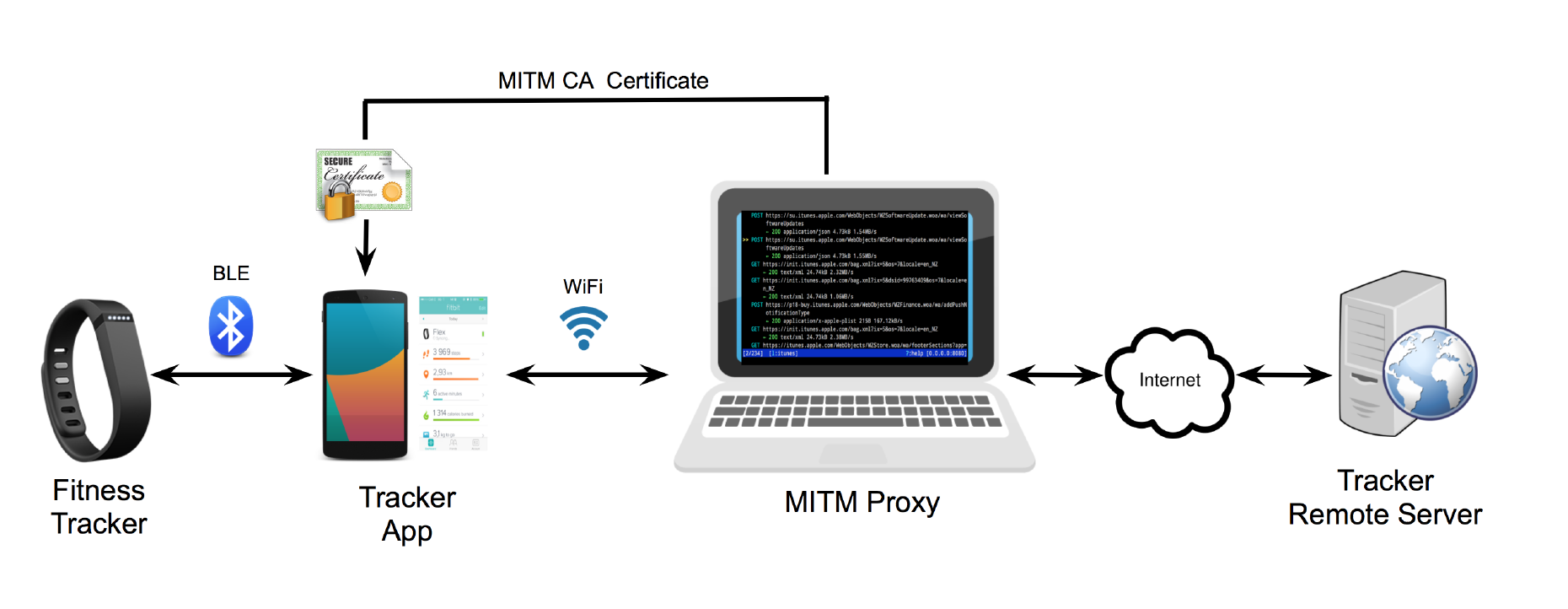}
  \vspace{-1.5em}
  \caption{Schematic illustration of the testbed used for protocol reverse engineering. Linux-based laptop used as wireless Internet gateway and running \gls{MITM} proxy.}
  \vspace{-1.5em}
  \label{fig:testbed}
\end{figure*}

\subsection{MITM Setup} 
To intercept the communication between the tracker and the remote server, we deploy an \gls{MITM} proxy on a Linux-based laptop acting as
a wireless Internet gateway, as illustrated in \autoref{fig:testbed}.  We install a fake CA certificate on an Android phone and
trigger tracker synchronization manually, using an unmodified Fitbit application.  The application synchronizes the tracker over Bluetooth LE
and forwards data between the tracker and the server over the Wi-Fi connection, encapsulating the information into JSON messages sent
over an HTTPS connection.  This procedure resembles typical user engagement with the tracker, however the MITM proxy allows us to intercept
all communications between the tracker and the server, as well as between the smartphone and the server.  In the absence of end-to-end encryption,
we can both capture and modify messages generated by the tracker.  Even with end-to-end encryption enabled, we can still
read the activity digests that the server provides to logged-in users, which are displayed by the app running on their smartphones.

\subsection{Wireshark Plugin Development and Packet Analysis}
To simplify the analysis process and ensure repeatability, we develop a custom frame dissector as stand-alone plugin programmed in C for the Wireshark network
analyzer~\cite{wireshark}.\footnote{The source code of our plug-in is available at \url{https://seemoo.de/fitbit-wireshark}.}
Developing this dissector involves cross-correlating the raw messages sent by the tracker with the server's JSON
responses to the client application.  After repeated experiments, we infer the many protocol fields that are present in
tracker-originated messages and that are encoded in different formats as detailed next.  We use the knowledge gained
to present these fields in a human-readable format in the protocol analyzer.

There are two types of tracker-originated messages we have observed during our analysis, which will be further described in the following sections:
\begin{enumerate}
  \item \textbf{Microdumps:} A summary of the tracker status and configuration.
  \item \textbf{Megadumps:} A summary of user activity data from the tracker.
\end{enumerate}

\begin{figure*}[ht]
\centering
  \includegraphics[width=0.8\linewidth]{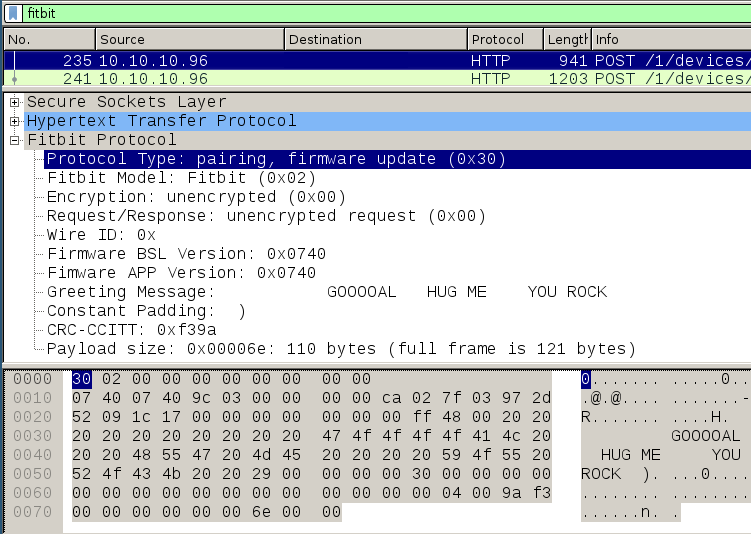}
  \caption{Generic microdump in plain-text, as displayed by the wireshark dissector we implement. Note the ability to filter by `fitbit' protocol type in the analyzer.}
  \label{fig:microdump-wireshark}
  \vspace{-1.5em}
\end{figure*}

\subsection{Microdump}
Depending on the action being performed by the user (e.g.\ authentication and pairing, synchronizing activity records),
the smartphone app makes HTTPS requests to the server using specific URLs, e.g.\ {\ttfamily POST \url{https://<fitbit\_server\_ip>/1/devices/client/.../validate.json?btle\_Name=Flex\&secret=null\&btAddress=}}\newline {\ttfamily\url{<6Byte\_tracker\_ID>}} for initial authentication.
Each basic action is accompanied by a so-called \textit{microdump}, which is required to identify the tracker,
and to obtain its state (e.g.\ its current firmware version).  Irrespective of whether or not the tracker implements protocol encryption,
the microdump header includes the tracker ID and firmware version, and is sent in plain-text.   \autoref{fig:microdump-wireshark}
illustrates a microdump sent along with a firmware update request, as interpreted by our Wireshark dissector.

We also note that the only validation feature that plain-text messages implement is a CRC-CCITT checksum, presumably used by the server
to detect data corruption in tracker-originated messages.  In particular, this acquired knowledge will allow us to inject generic messages
into the server and obtain replies, even when a valid tracker ID is already associated with a person's existing account.  Yet, microdumps only
contain generic information, which does not allow the spoofing of user activity records.  In what follows, we detail the format of messages
sent to the server to synchronize the tracked user activity.

Note that the plain-text format does not provide measures for verifying the integrity and authenticity of the message contents
except for a checksum, which is deterministically calculated from the values of the message fields.
This allows the adversary to inject generic messages to the server and receive replies, including information about whether a tracker ID is valid and
associated with a user account.

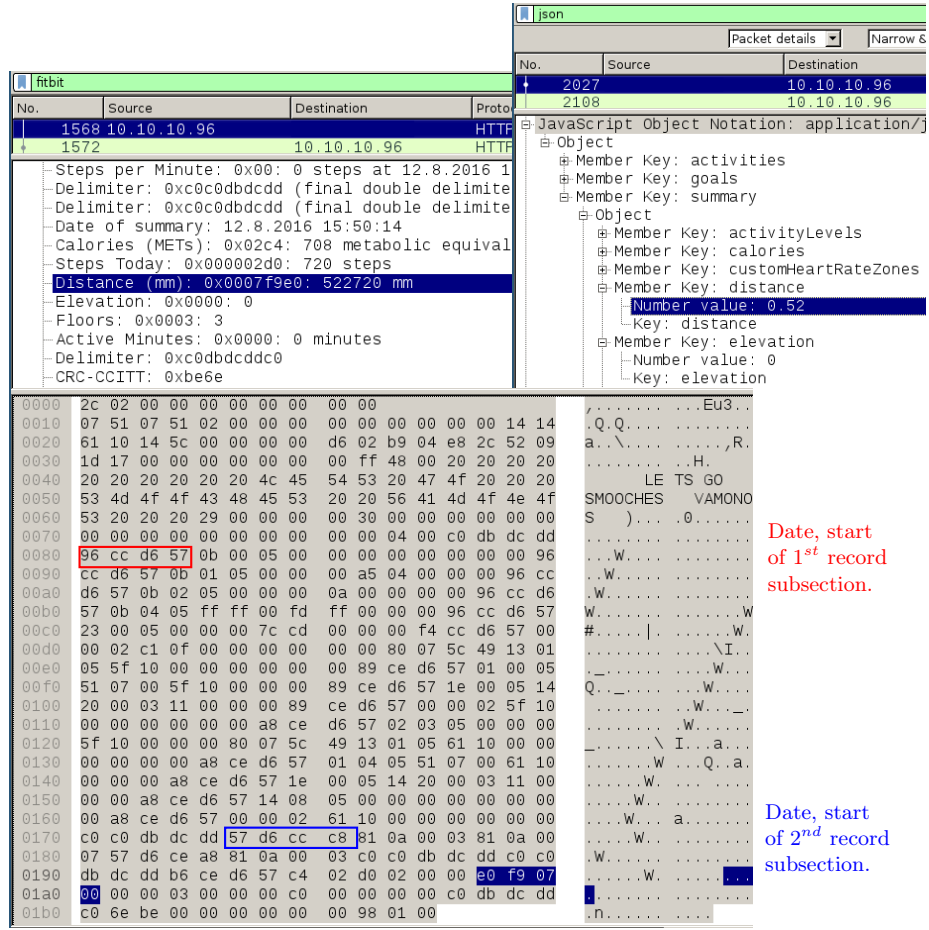
\begin{figure*}[!t]
  \centering
    \begin{tikzpicture}[thick,scale=0.9, every node/.style={scale=0.9}]
		\node[inner sep=0pt] (megadump) at (0,0) {\includegraphics[width=0.9\linewidth]{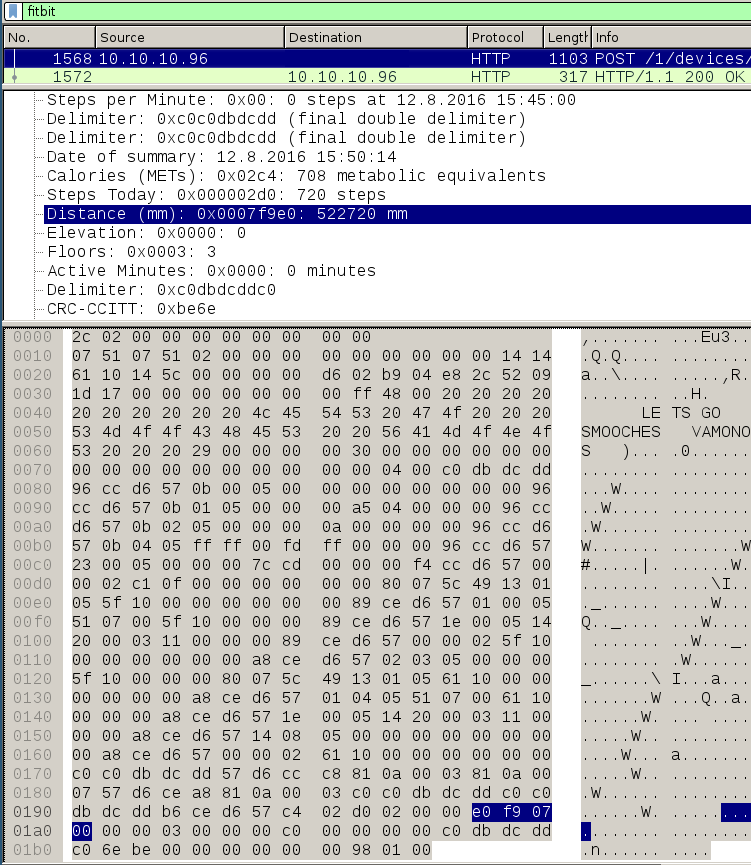}};
		\draw[-,thick,color=red] (-4.45,-0.7) rectangle (-2.8,-1.0);
		\draw[color=red, align=left] (6.6,-0.85) node {Date, start \\ of 1$^{st}$ record\\ subsection.};
		\draw[-,thick,color=blue] (-2.3,-4.85) rectangle (-0.35,-5.15);
		\draw[color=blue, align=left] (6.6,-5) node {Date, start \\ of 2$^{nd}$ record\\ subsection.};

		\node[inner sep=0pt] (json) at (5,4.5) {\includegraphics[width=0.5\linewidth]{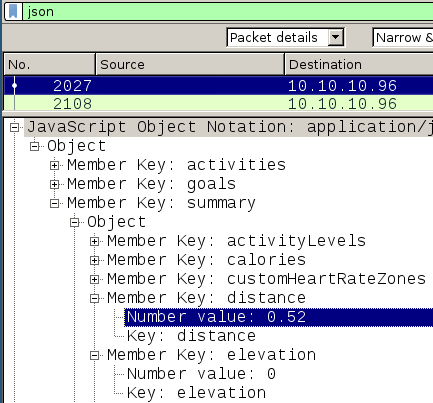}};		
	\end{tikzpicture}

    \caption{Megadump frame in plain-text format as transmitted to the Fitbit server (main window) and the human-readable JSON status response by the Fitbit Web API (top right).}
    \label{fig:megadump-json}
\end{figure*}


\subsection{Megadump Synchronization Message}

Step counts and other statistics are transmitted by the tracker in the form of a so-called \textit{megadump}.  Independent of encrypted or
plain-text mode, neither the Fitbit smartphone application nor the Galileo synchronization tool are aware of the exact meaning of
this payload.  The megadump is simply forwarded to the server, which in turn parses the message and responds with a reply.  This reply is
then forwarded (by the corresponding application) back to the tracker, confirming to the tracker that the data was synchronized
with the server successfully.

Despite this behavior, the Fitbit smartphone application---in contrast to Galileo---is aware of the user's statistics.  However,
this is due to the application making requests to the Fitbit Web API.  Once authenticated, this API can be used to retrieve
user information from the server in JSON format.  The Fitbit smartphone application
periodically synchronizes its display via the Fitbit Web API, allowing the user to see the latest information that was
uploaded by the most recent tracker megadump.  A plain-text example of this is shown in \autoref{fig:megadump-json}.  Note that the
Fitbit Web API separates data by type, such that not all information transmitted within one megadump is contained within one JSON
response.  From the megadump a total distance of \SI{522720}{\milli\meter} can be extracted,
which equals to the \SI{0.52}{\kilo\meter} from the JSON.

We use this information to reverse engineer and validate the megadump packet format, and have identified that each megadump
is split into the following sections: a header, one or more \textit{data sections}, and a footer.  These sections start with
a \textit{section start} sequence of bytes: \texttt{c0 cd db dc}; and end with a \textit{section terminator} byte: \texttt{c0}.
If the byte \texttt{c0} is required to be used within a data section, it is escaped in a manner similar to
RFC 1055.\footnote{A Non-standard for transmission of IP Data-grams over Serial Lines: SLIP}


\par
\begin{figure*}[t]
  \centering
  \includegraphics[width=\textwidth]{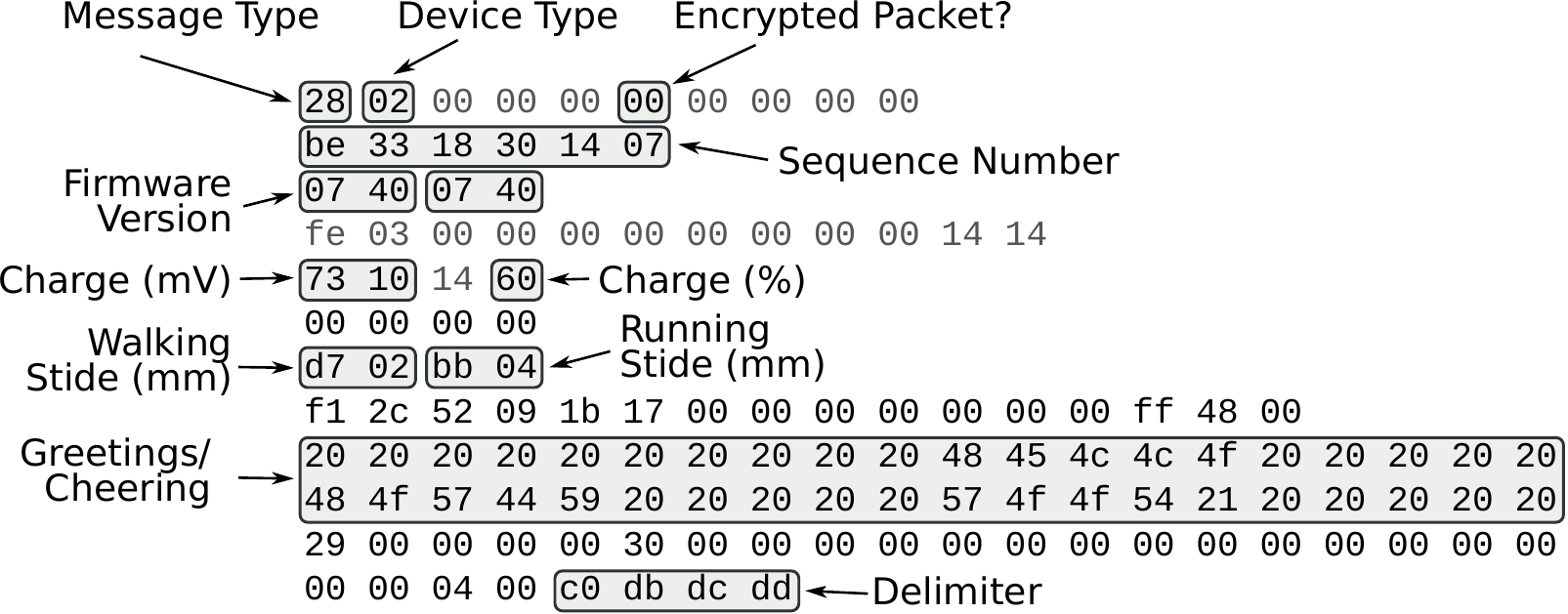}
  \caption{Megadump Header Structure}
  \label{fig:megadump-fields}
\end{figure*}

{\textbf{Message Header}}
The megadump header is very similar to the microdump header, but contains a few differences.  \autoref{fig:megadump-fields}
shows how this header is structured.

\par{\textbf{Data Sections}}
Following the header are one or more \textit{data sections}.  Each \textit{data section} contains various statistics in a particular format, and may
even be blank.  As previously mentioned, each data sections start with \texttt{c0 cd db dc}, and are terminated by a single \text{c0} character.
Therefore, the data sections are of variable length.  From the packets we have analyzed, it has been observed that there are typically four data sections, which all appear in the following order, and have the following format:

\par{(1) Daily Summary: }
The first data section contains activity information across a number of different absolute timestamps.  This section contains a series of fixed-length
records that begin with a little-endian timestamp, and end with a section terminator byte (\texttt{c0}).

\par
\begin{figure*}[t]
  \centering
  \includegraphics[width=0.5\textwidth]{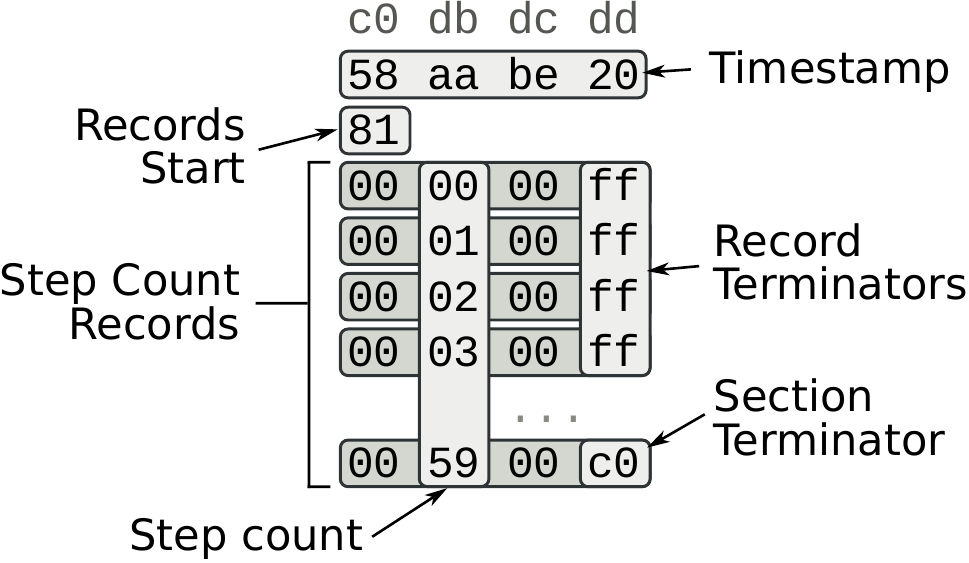}
  \caption{Per-minute Summary}
  \label{fig:megadump-per-minute}
\end{figure*}

{(2) Per-minute Summary: }
The next data section is a \textit{per-minute summary}, comprising a series of records that indicate user activity on a per-minute
granularity.  The structure of this data section is shown in \autoref{fig:megadump-per-minute}.

The section begins with a timestamp (unlike other timestamps, this field is big-endian), which acts as the \textit{base} time for this sequence
of step counts.  Each step count record is then an increment of a time period (typically two minutes), from this base time.  Following the timestamp 
is a byte indicating the start of the step count records.  The full meaning of this byte is unclear, but we believe it indicates the time period
between each step count record.  Following this, a series of records consisting of four bytes state the number of steps taken per-time period.
The second byte indicates the number of steps taken, and the fourth byte is either \texttt{ff} to indicate another record follows, or \texttt{c0}
(for the last record) to terminate the data section.

\par
\begin{figure*}[t]
  \centering
  \includegraphics[width=.7\textwidth]{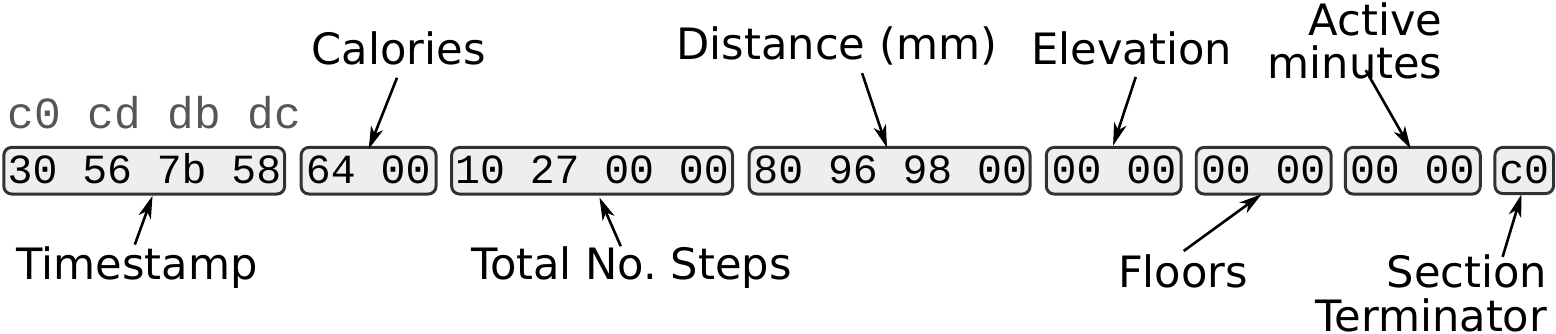}
  \caption{Megadump Summary Fields}
  \label{fig:megadump-summary}
\end{figure*}

{(3) Overall Summary: }
This data section contains a summary of the previous records, although as will be demonstrated later it is
not validated against ``per-minute'' or ``per-day'' data.  The format of this section is shown in \autoref{fig:megadump-summary}.

This section starts with a timestamp, indicating the base time for this summary data.  Following this timestamp is a 16-bit value
that holds the number of calories burned.  Following on from this is a 32-bit value containing the number of steps taken, and
a 32-bit value containing the distance travelled in millimeters.  Finally, the summary ends with elevation, floors climbed and
active minutes---all 16-bit values.

\par{(4) Alarms: }
The final data section contains information about what alarms are currently set on the tracker, and is typically empty
unless the user has instructed the tracker to create an alarm.

\par
\begin{figure*}[t]
  \centering
  \includegraphics[width=0.5\textwidth]{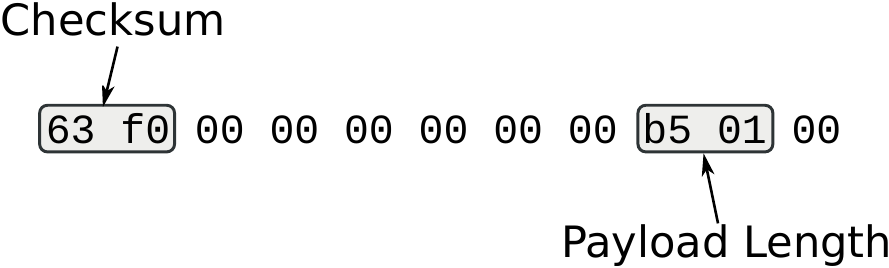}
  \caption{Megadump Footer Fields}
  \label{fig:megadump-footer}
\end{figure*}

{\textbf{Message Footer}}
The megadump footer contains a checksum and the size of the payload, as shown in \autoref{fig:megadump-footer}.

\section{Protocol-based Remote Spoofing}
\label{sec:protocol}
This section shows that the construction of a megadump packet containing fake information and the subsequent transmission to the Fitbit server is a viable approach for inserting fake step data into a user's exercise profile.  This attack does not actually require the possession of a physical tracker, but merely a known tracker ID to be associated with the user's Fitbit account. This means that one can fabricate fake data for any known and actively used tracker having a firmware version susceptible to this vulnerability. 
In order to construct a forged packet, however, the format of the message must be decoded and analyzed to determine the fields that must be populated.

\subsection{Submission of Fake Data}
\label{sec:sfd}
\afterpage{
\begin{figure}
  \centering
  \begin{minipage}{.45\textwidth}
    \includegraphics[height=2in]{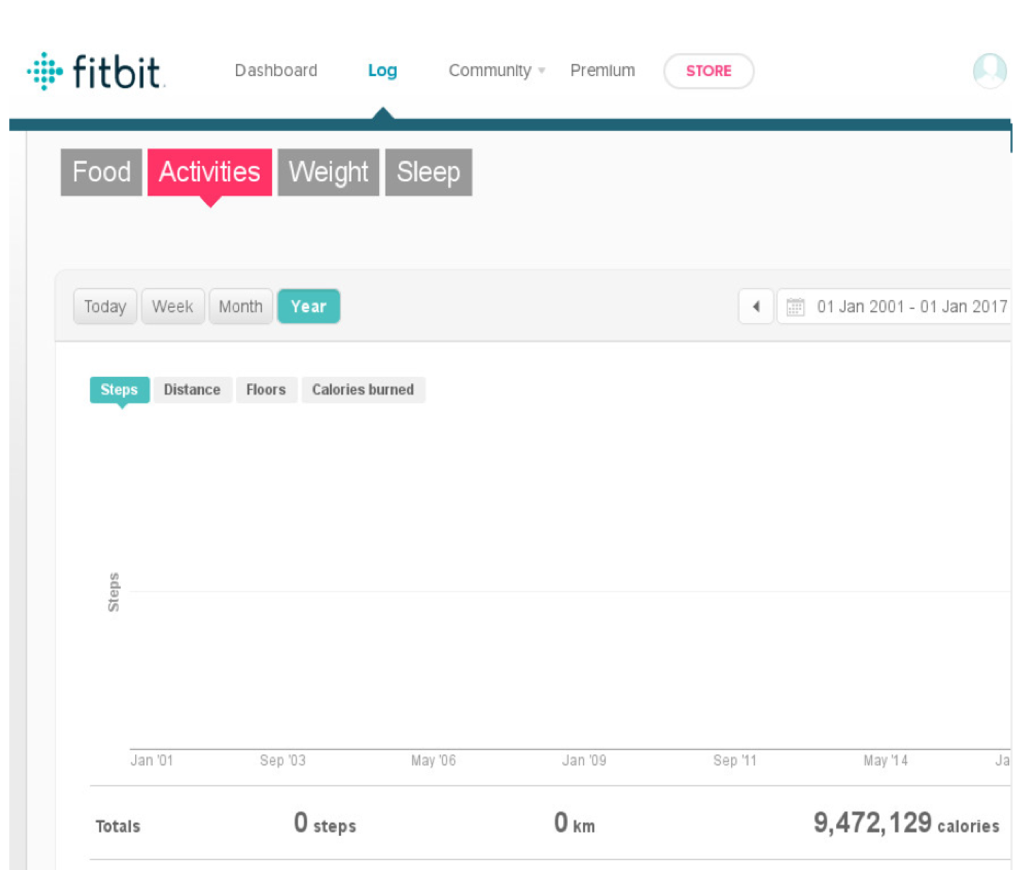}
    \subcaption{Before submission}
    \label{fig:di:before}
  \end{minipage}
  \qquad
  \begin{minipage}{.45\textwidth}\vspace{0.06in} 
    \includegraphics[height=1.95in]{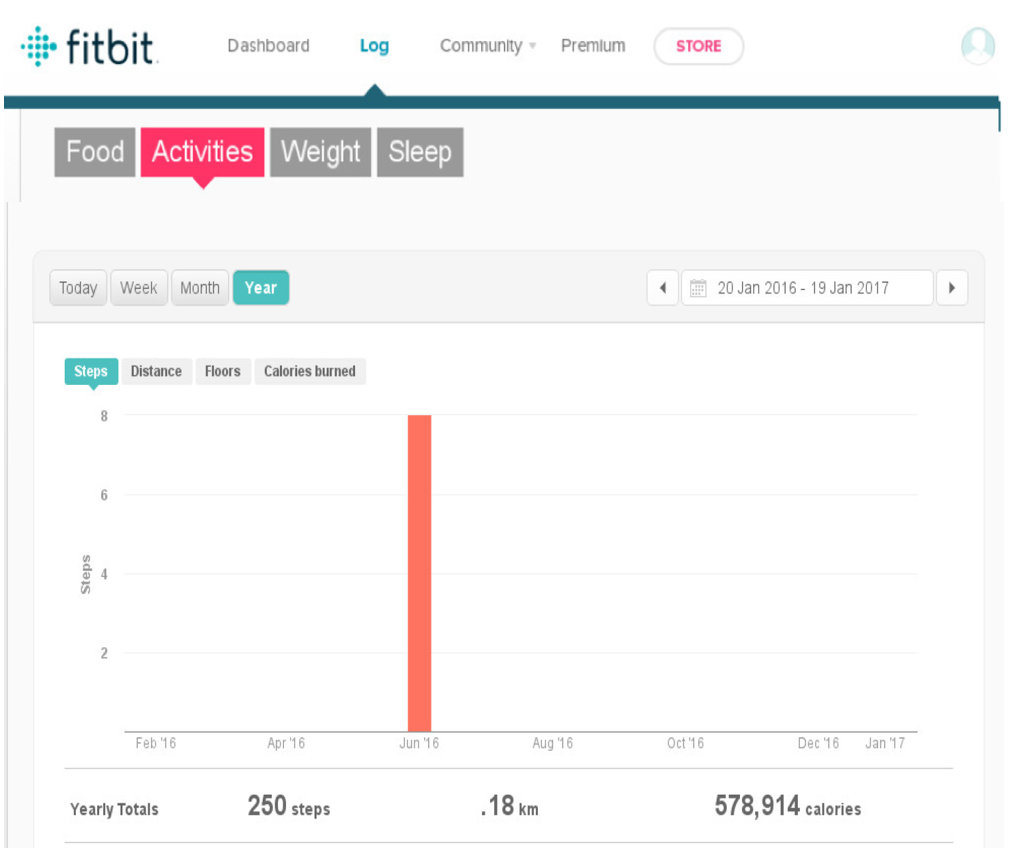}
    \subcaption{After submission}
    \label{fig:di:after}
  \end{minipage}
  \caption{The result of replaying data from another Fitbit tracker to a different tracker ID.
  \autoref{fig:di:before} shows the Fitbit user activity screen before the replay attack, and \autoref{fig:di:after} shows the results after the message is formed by changing the tracker ID, and
  submitted to the server.}
  \label{fig:direct-injection-replay}
\end{figure}
}

The Fitbit server has an HTTPS endpoint that accepts raw messages from trackers, wrapped in an XML description.  The raw message from the tracker is Base64 encoded, and contains various fields that describe the tracker's activity over a period of time.

The raw messages of the studied trackers may or may not be encrypted, but the remote server will accept either. Even though the encryption key for a particular tracker is unknown, it is possible to construct an unencrypted frame and submit it to the server for processing, associating it with an arbitrary tracker ID.  Provided that all of the fields in the payload are valid and the checksum is correct, the remote server will accept the payload and update
the activity log accordingly. In order to form such a message, the raw Fitbit frame must be Base64 encoded and placed within an XML wrapper
as shown in Listing~\ref{lst:lst1}: 

   \lstset{
    language=xml,
    tabsize=3,
    caption=Fitbit frame within an XML wrapper,
    label=lst:lst1,
    frame=shadowbox,
    rulesepcolor=\color{gray},
    xleftmargin=20pt,
    framexleftmargin=15pt,
    keywordstyle=\color{blue}\bf,
    commentstyle=\color{OliveGreen},
    stringstyle=\color{red},
    numbers=left,
    numberstyle=\tiny,
    numbersep=5pt,
    breaklines=true,
    showstringspaces=false,
    basicstyle=\scriptsize,
    emph={food,name,price},
    emphstyle={\color{magenta}},
    }
\begin{lstlisting}
<?xml version="1.0"?>
<galileo-client version="2.0">
 <client-info>
 <client-id>
    6de4df71-17f9-43ea-9854-67f842021e05
 </client-id>
 <client-version>1.0.0.2292</client-version>
 <client-mode>sync</client-mode>
 <dongle-version major="2" minor="5" />
 </client-info>
 <tracker tracker-id="F0609A12B0C0">
  <data>*** BASE64 PACKET DATA ***</data>
 </tracker>
</galileo-client>
\end{lstlisting}

The fabricated frame can be stored in a file, e.g.\ \texttt{payload}, and then submitted with the help of an HTTP \texttt{POST} request to the remote server as shown in Listing~\ref{lst:lst2}, after which the server will respond with
a confirmation message.  

 \lstset{
    language=bash,
    tabsize=3,
    caption=Submitting fake payload to the server ,
    label=lst:lst2,
    frame=shadowbox,
    rulesepcolor=\color{gray},
    xleftmargin=20pt,
    framexleftmargin=15pt,
    keywordstyle=\color{blue}\bf,
    commentstyle=\color{OliveGreen},
    stringstyle=\color{red},
    numbers=left,
    numberstyle=\tiny,
    numbersep=5pt,
    breaklines=true,
    showstringspaces=false,
    basicstyle=\scriptsize,
    emph={food,name,price},
    emphstyle={\color{magenta}},
    }
\begin{lstlisting}
$ curl -i -X POST https://client.fitbit.com/tracker/client/message 
-H "Content-Type: text/xml" 
--data-binary @payload
\end{lstlisting} 

\par{\textbf{Impersonation Attack: }}
 In order to test the susceptibility of the server to this attack, a frame from a particular tracker was
captured and re-submitted to the server with a \emph{different} tracker ID. The different tracker ID
was associated with a \emph{different} Fitbit user account. The remote server accepted the payload, and
updated the Fitbit user profile in question with identical information as for the genuine profile, confirming that simply
altering the tracker ID in the submission message allowed arbitrary unencrypted payloads to be accepted. \autoref{fig:direct-injection-replay} shows the Fitbit user activity logs before and after performing the impersonation attack. 
The fact that we are able to inject a data report associated to any of the studied trackers' IDs reveals both a severe DoS risk and the potential for a paid rogue service that would manipulate records on demand.
Specifically, an attacker could arbitrarily modify the activity records of random users, or manipulate the data recorded by the device of a target victim, as tracker IDs are listed on the packaging. Likewise, a selfish user may pay for a service that exploits this vulnerability to manipulate activity records on demand, and subsequently gain rewards.

\begin{figure*}
\vspace{-2em}
    \centering
  \begin{minipage}{.45\textwidth}
   \centering
    \includegraphics[height=2in]{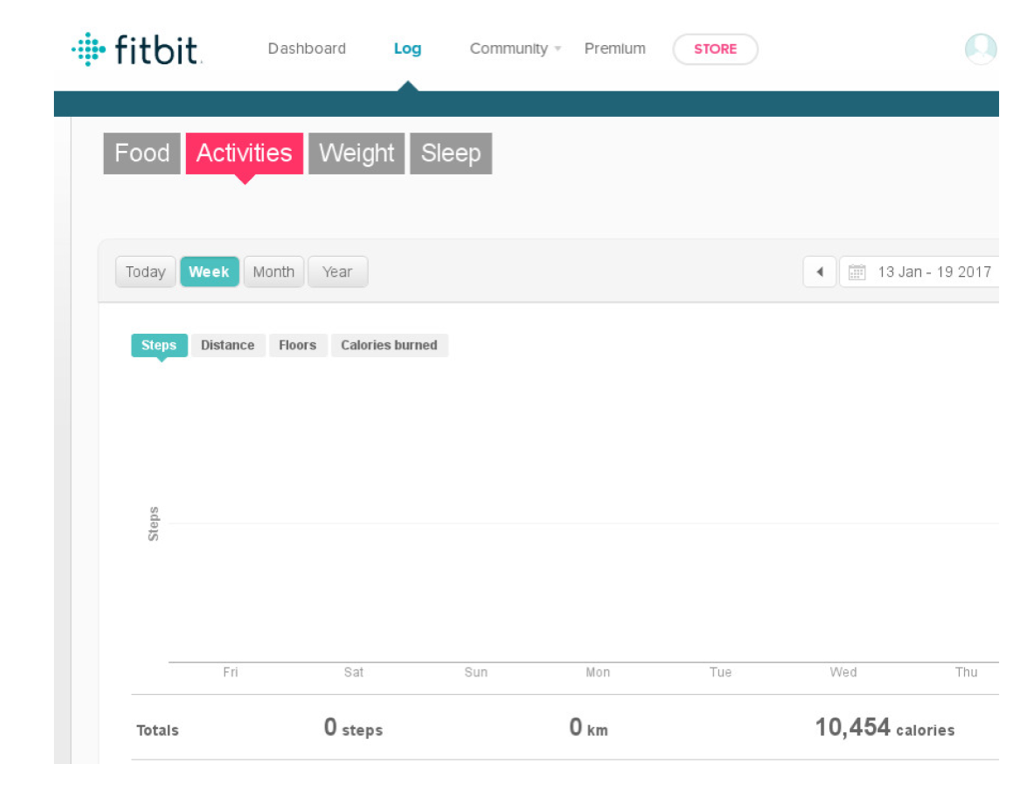}
    \subcaption{Before submission}
    \label{fig:co:before}
  \end{minipage}
  \qquad
  \begin{minipage}{.45\textwidth}
   \centering
    \includegraphics[height=2in]{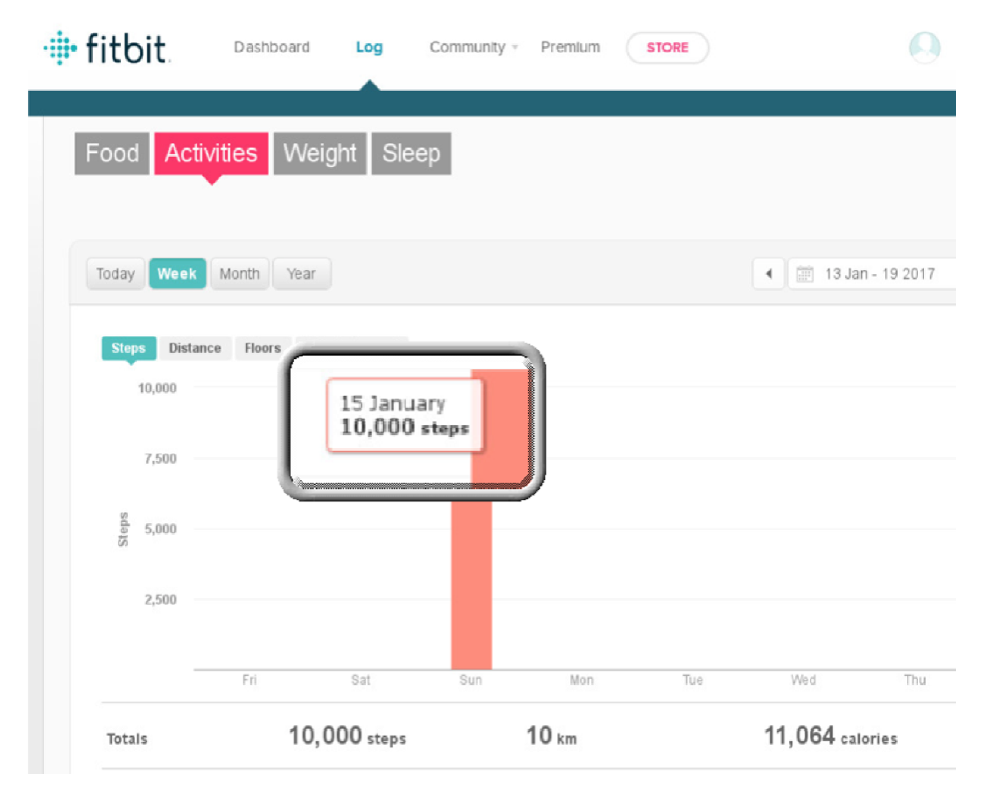}
    \subcaption{After submission}
    \label{fig:co:after}
  \end{minipage}
  \caption{\autoref{fig:co:before} shows the Fitbit user activity screen before fake data were
  submitted, and \autoref{fig:co:after} shows the screen after the attack.  In this example, 10000 steps and \SI{10}{km} were injected for the date of Sunday, January 15th, 2017 by fabricating a message containing the data shown in \autoref{fig:summary-section}.}
  \label{fig:direct-injection-specific}
\vspace{-2em}
\end{figure*}

\par{\textbf{Fabrication of Activity Data: }}
Using the information gained during the protocol analysis phase (see \autoref{sec:protocol-format}),
we constructed a message containing a frame with fake activity data and submitted it to the server, as discussed above.
To do this, the payload of a genuine message was used as a \textit{skeleton}, and each data section within the payload was cleared by removing all data bytes between the delimiters.  Then, the summary section was populated with fake data. Using only the summary section was enough to update the Fitbit user profile with fabricated step count and distance traveled information. The format of the summary section is shown in \autoref{fig:summary-section}, along with the fake
data used to form the fabricated message.

\begin{table}
\vspace{-2em}
\centering
\caption{Data inserted into the packet summary section}
\label{fig:summary-section}
\scriptsize
  \begin{tabular}{l|l|ll}
    \textbf{Range} & \textbf{Usage} & \textbf{Value} \\
    \hline
    \texttt{00-03} & Timestamp & \texttt{30 56 7b 58} & 15/01/17 \\
    \texttt{04-05} & Calories & \texttt{64 00} & 100 \\
    \texttt{06-09} & Number of Steps & \texttt{10 27 00 00} & 10000\\
    \texttt{0A-0D} & Distance in mm & \texttt{80 96 98 00} & 10000000 \\
    \texttt{0E-0F} & Elevation & \texttt{00 00 00 00} & 0
  \end{tabular} 
\end{table}

\autoref{fig:direct-injection-specific} again shows a before and after view of the Fitbit user activity screen, when the fake message is submitted.  In this example, the packet is constructed so that 10000 steps and a distance traveled of \SI{10}{km} were registered for the 15th of January 2017. This attack indicates that it is possible to create an arbitrary activity message and have the remote server accept it as a real update to the user's activity log.

\par{\textbf{Exploitation of Remote Server for Field Deduction: }}
A particular problem with the unencrypted packets was that it was not apparent how the value of the CRC field is calculated (unlike the CRC for encrypted packets).  However, if a message is sent to the server containing an invalid CRC, the server responds with a message containing information on what the correct CRC should be (see \autoref{lst:lst3}).

 \lstset{
    language=xml,
    tabsize=3,
    caption=Response from the Fitbit server when a payload with an invalid checksum is submitted.,
    label=lst:lst3,
    frame=shadowbox,
    rulesepcolor=\color{gray},
    xleftmargin=20pt,
    framexleftmargin=15pt,
    keywordstyle=\color{blue}\bf,
    commentstyle=\color{OliveGreen},
    stringstyle=\color{red},
    numbers=left,
    numberstyle=\tiny,
    numbersep=5pt,
    breaklines=true,
    showstringspaces=false,
    basicstyle=\scriptsize,
    emph={food,name,price},
    emphstyle={\color{magenta}},
    }
\begin{lstlisting}
$ curl -i -X POST <target-url> --data-binary @payload
<?xml version="1.0" encoding="UTF-8" standalone="yes"?>
<galileo-server version="2.0">
  <error>INVALID_DEVICE_DATA:com.fitbit.protocol.serializer.DataProcessingException: Parsing field
  [signature] of the object of type CHECKSUM. IO error -&gt; Remote checksum [2246|0x8c6] and local
  checksum [60441|0xec19] do not match.</error>
</galileo-server>
\end{lstlisting}

This information can be used to reconstruct the packet with a valid CRC. Such an exploit must be used sparingly, however, as the remote server will refuse to process further messages if an error threshold is met, until a lengthy timeout (on the order of hours) expires.


\section{Hardware-Based Local Spoofing}
\label{sec:hardware}

\afterpage{
\begin{figure}
\centering
  \includegraphics[width=1\linewidth]{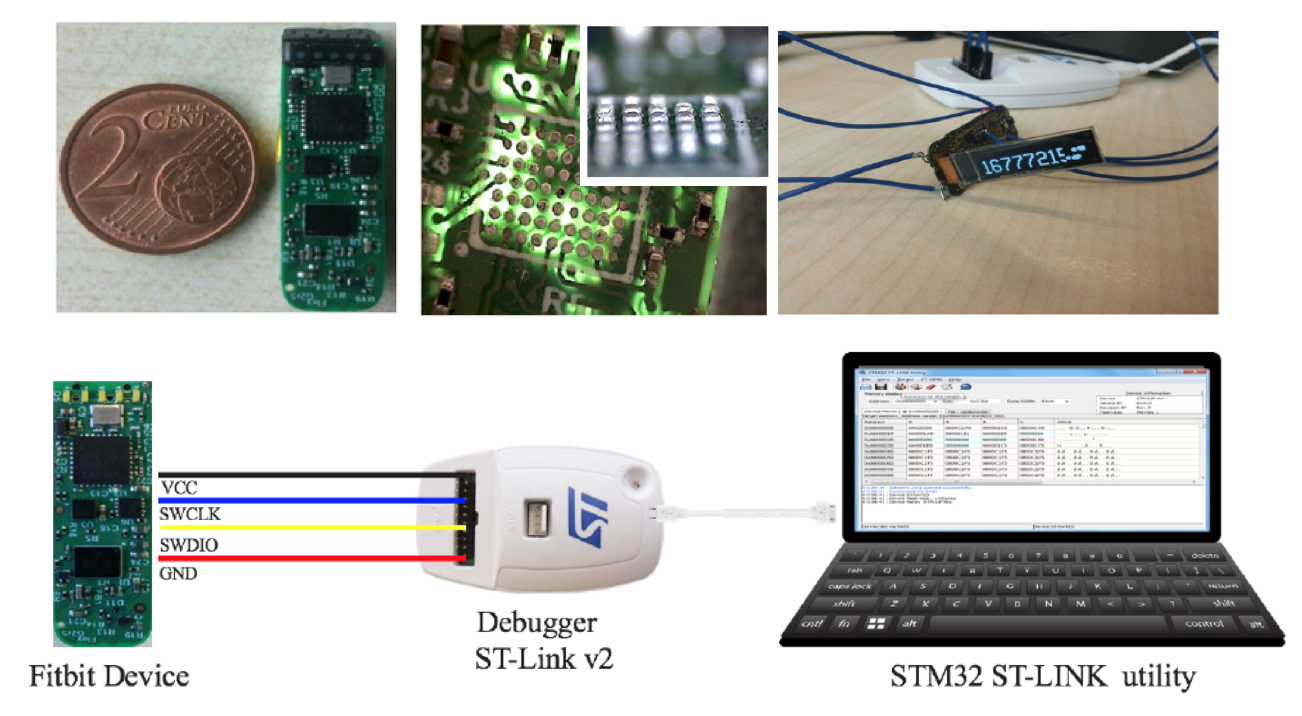}
  \caption{Fitbit tear-down and connecting Fitbit micro-controller to the debugger.}
  \label{fig:tear1}
\end{figure}
}

We now demonstrate the feasibility of hardware-based spoofing attacks focusing on Fitbit Flex and Fitbit One devices. We first conducted an analysis of the Fitbit protocol as previously described in \autoref{sec:protocol-format}. However, since the newest firmware (Fitbit 7.81) uses end-to-end encryption with a device-specific key, the data cannot be manipulated using \gls{MITM} attacks, as described in the previous section.
Therefore, we resort to a physical attack on the tracker's hardware. We reverse engineered the hardware layout of the devices to gain memory access, which enabled us to inject arbitrary stepcount values into memory, which the tracker would send as valid encrypted frames to the server.

\subsection{Device Tear-Down}
In order to understand how to perform the hardware attack, we needed to tear down the devices. In the following section, we give an overview of the tools required for this process.
\par{\textbf{Tools: }}
The tools to perform the hardware attack were relatively inexpensive and easy to purchase.  To accomplish the attack, we used
\begin{enumerate*}[label=(\roman*)]
\item a digital multimeter,
\item a soldering iron, thin gauge wire, flux
\item tweezers,
\item a soldering heat gun,
\item the ST-LINK/v2 in circuit debugger/programmer, and
\item the STM32 ST-LINK utility. 
\end{enumerate*}

The digital multimeter was used to locate the testing pins associated with the debug interface of the microcontroller. However, attackers performing the attack would not require a multimeter, as long as the layout of the testing pins is known. The soldering heat gun and tweezers were utilized to perform the mechanical tear-down of the device casing. The soldering iron and accessories were used to solder wires to the identified testing pins. We used the ST-LINK/v2 and STM32 ST-LINK utilities to connect to the device in order to obtain access to the device's memory.

 \par{\textbf{Costs:  }} 
The required tools for performing the hardware attack are relatively cheap. The STLINK/v2 is a small debugger/programmer that connects to the PC using a common mini-USB lead and costs around \$15. The corresponding STM32 ST-LINK utility is a full-featured software interface for programming STM32 microcontrollers, using a mini-USB lead. This is free Windows software and that can be downloaded from ST\footnote{\url{http://www.st.com/en/embedded-software/stsw-link004.html}}. General-purpose tools (e.g. hair dryer) can be employed to tear-down the casing. Therefore the total costs make the attack accessible to anyone who can afford a fitness tracker. We argue that hardware modifications could also be performed by a third party in exchange of a small fee, when the end user lacks the skills and/or tools to exploit hardware weaknesses in order to obtain financial gains.


 \par{\textbf{Tear-Down Findings: }}
According to our tear-down of the Fitbit trackers (Fitbit Flex and Fitbit One), as shown in \autoref{fig:tear1}, the main chip on the motherboard is an ARM Cortex-M3 processor.  This processor is an ultra-low-power 32-bit MCU, with different memory banks such as 256KB flash, 32KB SRAM and 8KB EEPROM. The chip used for Fitbit Flex is \emph{STM32L151UC WLCSP63} and for Fitbit One \emph{STM32L152VC UFBGA100}. The package technology used in both micro-controllers is \gls{BGA} which is a surface-mount package with no leads and a grid array of solder balls underneath the integrated circuit. Since the required specifications of the micro-controller used in Fitbit trackers are freely available, we were able to perform hardware reverse-engineering (\gls{RE}).

%
\subsection{Hardware RE to Hunt Debug Ports}
We discovered a number of testing points at the back of the device's main board.  Our main goal was to identify the testing points connected to debug interfaces. According to the IC's datasheet, there are two debug interfaces available for \emph{STM32L}:
\begin{enumerate*}[label=(\roman*)]
\item \gls{SWD} and
\item \gls{JTAG}
\end{enumerate*}.

\begin{figure}

\begin{minipage}{\textwidth}

\begin{subfigure}{\linewidth}
\centering
\begin{tabular}{|l|c|c|}
\hline
\textbf{ST-LINK/V2 } & \textbf{SWD Pins} & \textbf{Description} \\ \hline
Pin 1           &  Vcc                 & Target board Vcc                \\ \hline
Pin 7           &  SWDIO              & The SWD Data Signal                \\ \hline
Pin 8             & GND                & Ground               \\ \hline
Pin 9            & SWCLK                 & The SWD Clock Signal              \\ \hline
Pin 15             & RESET                 &  System Reset             \\ \hline
\end{tabular}

\end{subfigure}

\end{minipage}
\caption{Connecting the tracker to the debugger.}  \label{fig:pinheader}
\end{figure}

We found that the Fitbit trackers were using the \gls{SWD} interface. However, the \gls{SWD} pins were obfuscated by placing them among several other testing points without the silkscreen identifying them as testing points. \gls{SWD} technology provides a 2-pin debug port, a low pin count and high-performance alternative to \gls{JTAG}. The \gls{SWD} replaces the \gls{JTAG} port with a clock and single bidirectional data pin, providing test functionality and real-time access to system memory. We selected a straightforward approach to find the debug ports (other tools that can be exploited include \emph{Arduino+JTAGEnum} and \emph{Jtagulator}). We removed the micro-controller from the device \glspl{PCB}. Afterward, using the IC's datasheet and a multimeter with continuity tester functionality, we traced the debug ports on the device board, identifying the testing points connected to them. 

\subsection{Connecting Devices to the Debugger}
After discovering the \gls{SWD} debug pins and their location on the \gls{PCB},  we soldered wires to the debug pins.  We connected the debug ports to ST-LINK v2 pin header, according to \autoref{fig:pinheader}.

\par{\textbf{Dumping the Firmware: }}
After connecting to the device micro-controller, we were able to communicate with MCU as shown in \autoref{fig:tear1}. We extracted the entire firmware image since memory readout protection was not activated. There are three levels of memory protection in the STM32L micro-controller:
\begin{enumerate*}[label=(\roman*)]
\item level 0: \emph{no readout protection}, 
\item level 1:  \emph{memory readout protection}, the Flash memory cannot be read from or written to, and
\item level 2: \emph{chip readout protection}, debug features and boot in RAM selection are disabled (JTAG fuse). 
\end{enumerate*}
 We discovered that in the Fitbit Flex and the Fitbit One, memory protection was set to \emph{level 0}, which means there is no memory readout protection. This enabled us to extract the contents of the different memory banks (e.g., FLASH, SRAM, ROM, EEPROM) for further analysis. 

Note that it is also possible to extract the complete firmware via the \gls{MITM} setup during an upgrade process (if the tracker firmware does not use encryption). In general, sniffing is easier to perform, but does not reveal the memory layout and temporal storage contents. Moreover, hardware access allows us to change memory contents at runtime.

\par{\textbf{Device Key Extraction: }}
We initially sniffed communications between the Fitbit tracker and the Fitbit server to see whether a key exchange protocol is performed, which was not the case. Therefore, we expected pre-shared keys on the Fitbit trackers we connected to, including two different Fitbit One and three different Fitbit Flex devices.  We read out their EEPROM and discovered that the device encryption key is stored in their EEPROM. Exploring the memory content, we found the exact memory addresses where the 6-byte serial ID and 16-byte encryption key are stored, 
as shown in \autoref{fig:key}.  We confirm that each device has a \emph{device-specific key} which likely is programmed into the device during manufacturing~\cite{Schellevis}.

\par{\textbf{Disabling the Device Encryption: }}
By analyzing the device memory content, we discovered that by flipping one byte at a particular address in EEPROM, we were able to force the tracker to operate in unencrypted mode and disable the encryption. Even trackers previously communicating in encrypted mode switched to plaintext after modifying the encryption flag (byte). \autoref{fig:key} illustrates how to flip the byte, such that the the tracker sends all sync messages in plaintext format (Base64 encoded) disabling encryption.

\afterpage{
\begin{figure}[H]
\centering
  \includegraphics[height=1.75in]{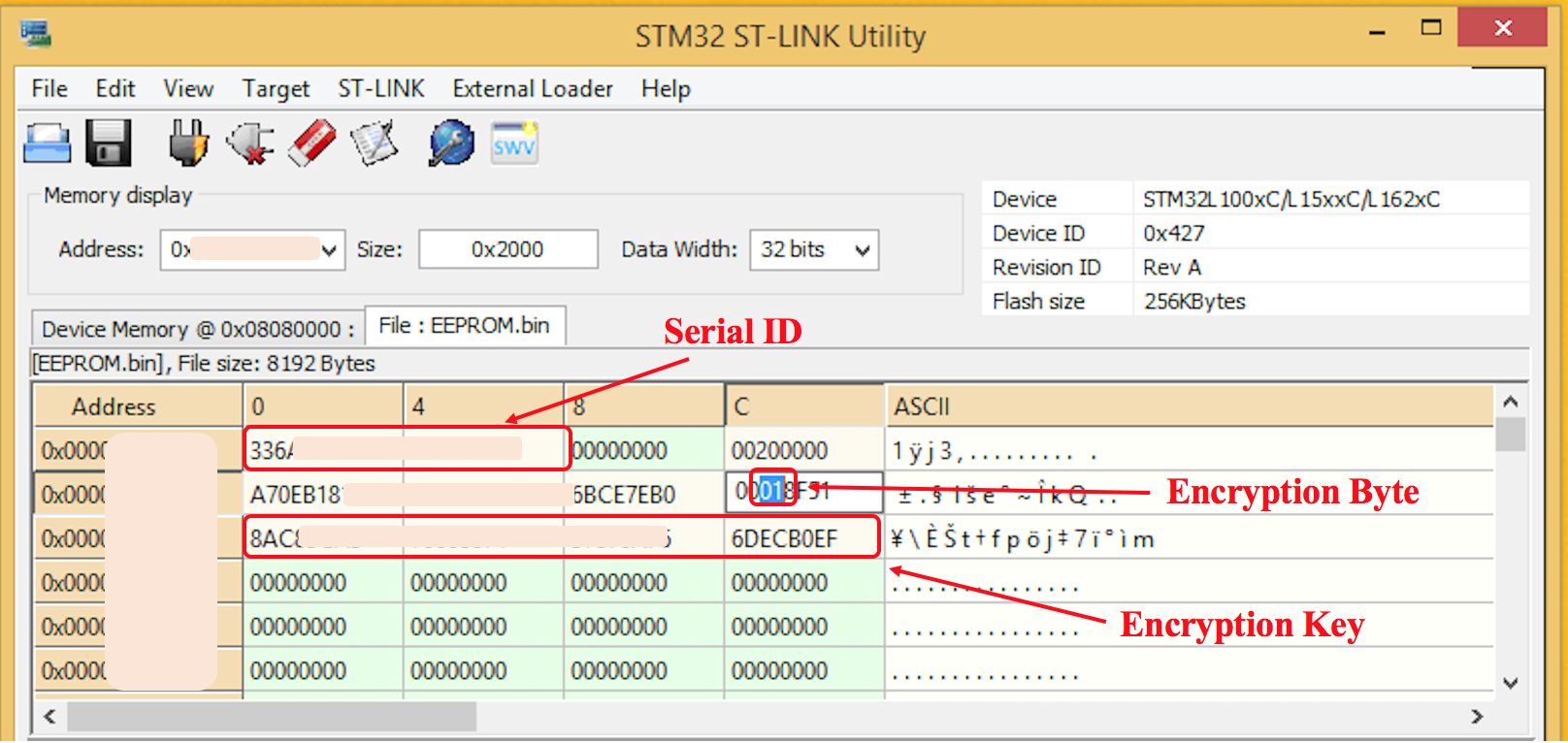}
  \caption{Device key extraction and disabling encryption.}
  \label{fig:key}
\end{figure}

}

\par{\textbf{Injecting Fabricated Data Activities: }}
We investigated the EEPROM and SRAM content to find the exact memory addresses where the total step count and other data fields are stored. Based on our packet format knowledge and previously sniffed megadumps, we found that the activity records were stored in the EEPROM in the same format. Even encrypted frames are generated based on the EEPROM plaintext records. Therefore, oblivious falsified data can be injected, even with the newest firmware having encryption enabled.  As it can be seen in \autoref{fig:fitapp} and \autoref{fig:web}, we managed to successfully inject \texttt{0X00FFFFFF} steps equal to \num{16777215} in decimal into Fitbit server by modifying the corresponding address field in the EEPROM and subsequently synchronising the tracker with the server. 

\afterpage{%
\begin{figure}[H]
\centering
\begin{minipage}{.4\textwidth}

\centering
\includegraphics[height=2.1in]{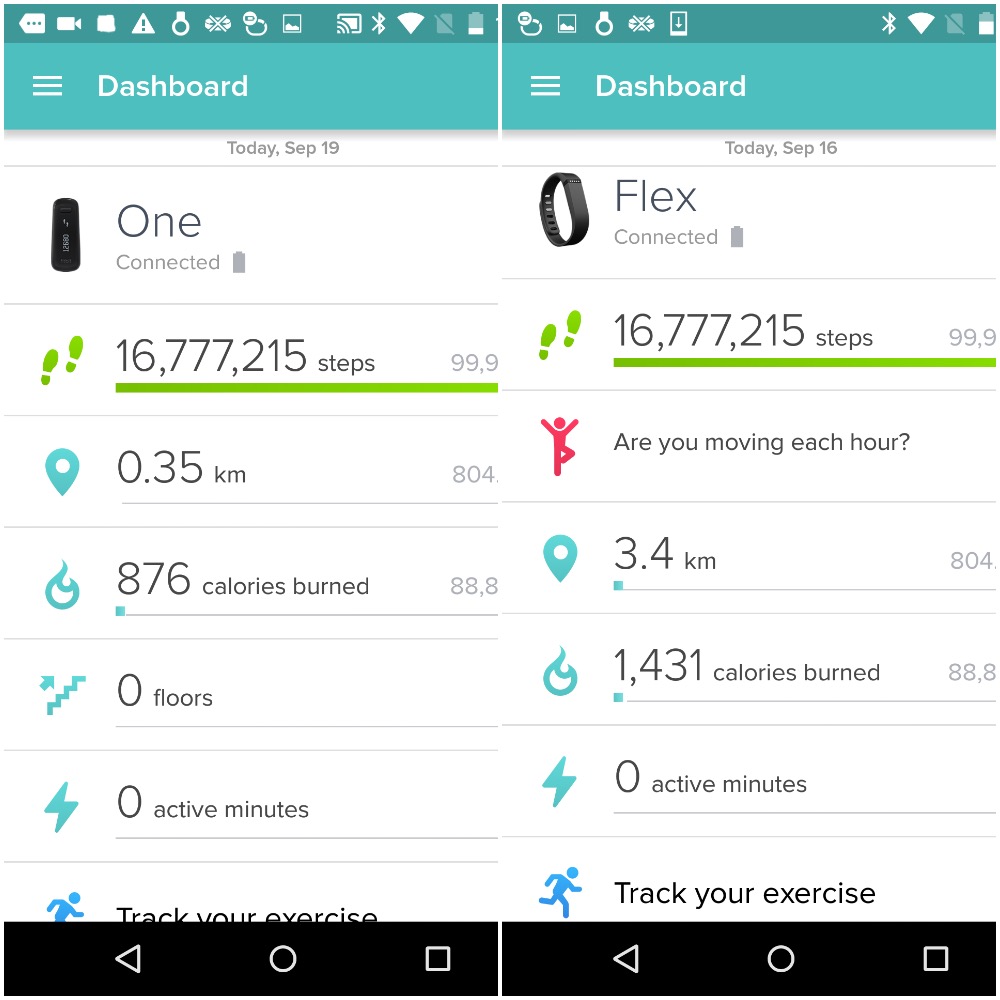}
\subcaption{Fitbit app}
 \label{fig:fitapp}
\end{minipage}
 \qquad
 \begin{minipage}{.50\textwidth}
\centering
\includegraphics[height=2.1in]{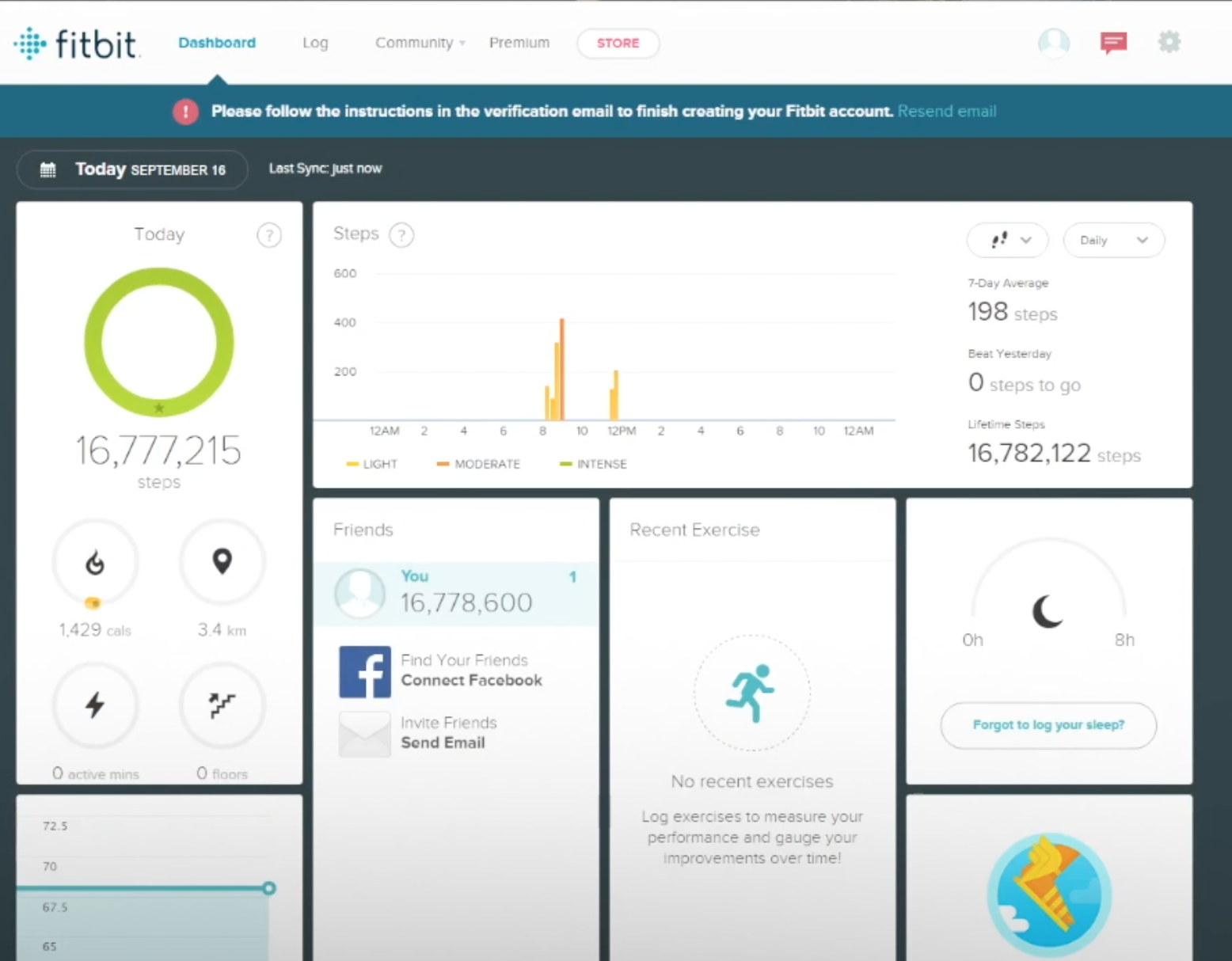}
\subcaption{Fitbit web interface}
 \label{fig:web}
\end{minipage}
 \caption{The results of injecting fabricated data.  \autoref{fig:fitapp} shows the Fitbit app screenshot, and \autoref{fig:web} demonstrates the Fitbit web interface.}
\end{figure}
}


\section{Discussion}

In this section we give a set of implementation guidelines for fitness trackers. While Fitbit is currently the only manufacturer that puts effort into securing trackers~\cite{Fereidooni:2017}, our guidelines also apply to other health-related IoT devices. We intend to transfer the lessons learned into open security and privacy standards that are being developed.\footnote{See https://www.thedigitalstandard.org}

False data injection as described in the previous sections is made possible by a combination of some of the design choices in the implementation of the Fitbit trackers and in the communication protocol utilized between the trackers and Fitbit application servers. These design choices relate to how encryption techniques have been applied, the design of the protocol messages, and the implementation of the hardware itself.  To overcome such weaknesses in future system designs, we propose the following mitigation techniques.
\par{\textbf{Application of encryption techniques: }}
The examined trackers support full end-to-end encryption, but do not enforce its use consistently.\footnote{During discussions we had with Fitbit, the company stressed that models launched after 2015 consistently enforce encryption in the communications between the tracker and server.} This allows us to perform an in-depth analysis of the data synchronization protocol and ultimately fabricate messages with false activity data, which were
accepted as genuine by the Fitbit servers.

\vspace{-.1cm}
\begin{suggestion} 
	End-to-end encryption between trackers and remote servers should be consistently enforced, if supported by device firmware.
\end{suggestion}

\par{\textbf{Protocol message design: } }
Generating valid protocol messages (without a clear understanding of the CRC in use) is enabled by the fact that the server
responds to invalid messages with information about the expected CRC values, instead of a simple ``invalid CRC'', or
a more general ``invalid message'' response.

\vspace{-.1cm}
\begin{suggestion}
	Error and status notifications should not include additional information related to the contents of actual protocol messages. 
\end{suggestion}

\vspace{-.2cm}
CRCs do not protect against message forgery, once the scheme is known. For authentication, there is already a scheme in place to
generate subkeys from the device key~\cite{Schellevis}. Such a key could also be used for message protection.

\vspace{-.1cm}
\begin{suggestion}
	Messages should be signed with an individual signature subkey which is derived from the device key.
\end{suggestion}

\vspace{-.2cm}
\par{\textbf{Hardware implementation: }}
The microcontroller hardware used by both analyzed trackers provides memory readout protection mechanisms, but were not
enabled in the analyzed devices. This opens an attack vector for gaining access to tracker memory and allows us to
circumvent even the relatively robust protection provided by end-to-end message encryption as we were able to modify
activity data directly in the tracker memory. Since reproducing such hardware attacks given the necessary background
information is not particularly expensive, the available hardware-supported memory protection measures should be applied
by default.

\vspace{-.1cm}
\begin{suggestion} 
	Hardware-supported memory readout protection should be applied.
\end{suggestion}
\vspace{-.3cm}
Specifically, on the MCUs of the investigated tracking devices, the memory of the hardware should be protected by enabling
chip readout protection level 2.

\par{\textbf{Fraud detection measures: }}
In our experiments we were able to inject fabricated activity data with clearly unreasonably high performance values
(e.g.\ more than 16 million steps during a single day). This suggests that data should be monitored more closely by the
servers before accepting activity updates.

\vspace{-.1cm}
\begin{suggestion}
	Fraud detection measures should be applied in order to screen for data resulting from malicious modifications or
	malfunctioning hardware.
\end{suggestion}

\vspace{-.3cm}
For example, accounts with unusual or abnormal activity profiles should be flagged and potentially disqualified, if
obvious irregularities are detected. \newline


\section{Related Work}
\label{sec:related}

Researchers at the University of Toronto~\cite{Andrew} have investigated transmission security, data integrity, and Bluetooth privacy of eight fitness trackers including Fitbit Charge HR. They focused on transmission security, specifically at whether or not personal data is encrypted when transmitted over the Internet in order to protect confidentiality.  They also examined data integrity concentrating on whether or not fitness data can be considered authentic records of activity that have not been tampered with. They did not attempt to reverse engineer the proprietary encoding or encryption used for transmitting data. 

In 2013, Rahman \emph{et al.}~\cite{Rahman:2013} studied the communication between Fitbit Ultra and its base station as well as the associated web servers. According to Rahman \emph{et al.}, Fitbit users could readily upload sensor data from their Fitbit device onto the web server, which could then be viewed by others online. They observed two critical vulnerabilities in the communication between the Fitbit device's base station, and the web server. They claimed that these vulnerabilities could be used to violate the security and privacy of the user. Specifically, the identified vulnerabilities consisted of the use of plaintext login information and plaintext HTTP data processing. Rahman \emph{et al.} then proposed FitLock as a solution to the identified vulnerabilities. These vulnerabilities have been patched by Fitbit and no longer exist on contemporary Fitbit devices. 
Zhou \emph{et al.}~\cite{Zhou} followed up on Rahman's work by identifying shortcomings in their proposed approach named FitLock, but did not mention countermeasures to mitigate the vulnerabilities that they found. In 2014, Rahman \emph{et al.} published another paper detailing weaknesses in Fitbit's communication protocol, enabling them to inject falsified data to both the remote web server and the fitness tracker. The authors proposed SensCrypt, a protocol for securing and managing low power fitness trackers~\cite{Rahman:2016}.
Note that Fitbit's communication paradigm has changed considerably since Fitbit Ultra, which uses ANT instead of Bluetooth, and is not supported by smartphone applications, but only by a Windows program last updated in 2013. Neither the ANT-based firewalls FitLock nor SensCrypt would work on recent Fitbit devices. Transferring their concept to a Bluetooth-based firewall would not help against the attacks demonstrated in this paper, since hardware attacks are one level below such firewalls, while our protocol attacks directly target the Fitbit servers.

Cyr \emph{et al.}~\cite{Britt} analyzed the Fitbit Flex ecosystem. They attempted to do a hardware analysis of the Fitbit device but because of the difficulties associated with debugging the device they decided to focus on other parts such as Bluetooth LE, the associated Android app and network analysis. The authors explained the data collected by Fitbit from its users, the data Fitbit provided to Fitbit users, and methods of recovering data not made available to device owners. 

In the report released by AV TEST~\cite{Eric:2015}, the authors tested nine fitness trackers including Fitbit Charge and evaluated their security and privacy. The authors tried to find out how easy it is to get the fitness data from the fitness band through Bluetooth or by sniffing the connection to the cloud during the synchronization process. 

AV TEST reported some security issues in Fitbit Charge \cite{Eric:2016}. They discovered that Fitbit Charge with firmware version 106 and lower allows non-authenticated smartphones to be treated as authenticated if an authenticated smartphone is in range or has been in range recently. Also, the firmware version allowed attackers to replay the tracker synchronization process. 
Both issues have been now fixed by Fitbit.

 In~\cite{Schellevis}, the authors captured the firmware image of the Fitbit Charge HR during a firmware update. 
They reversed engineer the cryptographic primitives used by the Fitbit Charge HR activity tracker and recovered the authentication protocol. Moreover, they obtained the cryptographic key that is used in the authentication protocol from the Fitbit Android application. The authors found a backdoor in previous firmware versions and exploiting this backdoor they extracted the device specific encryption key from the memory of the tracker using Bluetooth interface.  Memory readout has been fixed in recent firmware versions.

Principled understanding of the Fitbit protocol remains open to investigation as the open-source community continues to reverse-engineer message semantics and server responses~\cite{Galileo}.


\section{Conclusion}
\label{sec:conclusion}

Trusting the authenticity and integrity of the data that fitness trackers generate is paramount, as the records
they collect are being increasingly utilized as evidence in critical scenarios such as court trials and the adjustment
of healthcare insurance premiums. In this paper, we conducted an in-depth security analysis of two models of popular
activity trackers commercialized by \texttt{Fitbit}, the market leader, and we revealed serious security and privacy
vulnerabilities present in these devices. Additionally, we reverse engineered the primitives governing the communication
between these devices and cloud-based services, implemented an open-source tool to extract sensitive personal
information in human-readable format and demonstrated that malicious users could inject spoofed activity records to
obtain personal benefits. To circumvent the end-to-end protocol encryption mechanism present on the latest firmware,
we performed hardware-based \gls{RE} and documented successful injection of falsified data that appears legitimate to
the Fitbit cloud.
We believe more rigorous security controls should be enforced by manufacturers to verify the authenticity of fitness data. To this end, we provided a set of guidelines to be followed to address the vulnerabilities identified.


\begin{thebibliography}{20}
\bibitem{forbes:2016}
{Forbes}.
\newblock Wearable tech market to be worth \$34 billion by 2020.
\newblock   \url{https://www.forbes.com/sites/paullamkin/2016/02/17/wearable-tech-market-to-be-worth-34-billion-by-2020},   February 2016.

\bibitem{idc:2017}
{International Data Corporation}.
\newblock Worldwide quarterly wearable device tracker.
\newblock \url{https://www.idc.com/tracker/showproductinfo.jsp?prod_id=962},  March 2017.

\bibitem{mashable}
{Mashable}.
\newblock {Husband learns wife is pregnant from her Fitbit data}.
\newblock \url{http://mashable.com/2016/02/10/fitbit-pregnant/}, Feb. 2016.

\bibitem{wsj:2016}
{The Wall Street Journal}.
\newblock Prosecutors say {Fitbit} device exposed fibbing in rape case.
\newblock   \url{http://blogs.wsj.com/law/2016/04/21/prosecutors-say-fitbit-device-exposed-fibbing-in-rape-case/},  April 2016.

\bibitem{guardian:2014}
{The Guardian}.
\newblock {Court sets legal precedent with evidence from Fitbit health tracker}.
\newblock   \url{https://www.theguardian.com/technology/2014/nov/18/court-accepts-data-fitbit-health-tracker},  November 2014.

\bibitem{vitalityhealth}
{VitalityHealth}.
\newblock \url{https://www.vitality.co.uk/rewards/partners/activity-tracking/}.

\bibitem{achievemint}
{AchieveMint}.
\newblock \url{https://www.achievemint.com}.

\bibitem{stepbet}
{StepBet}.
\newblock \url{https://www.stepbet.com/}.


\bibitem{Rahman:2013}
Mahmudur Rahman, Bogdan Carbunar, and Madhusudan Banik.
\newblock {Fit and Vulnerable: Attacks and Defenses for a Health Monitoring Device}.
\newblock In {\em Proc. Privacy Enhancing Technologies Symposium (PETS)},  Bloomington, IN, USA, July 2013.

\bibitem{Britt}
Britt Cyr, Webb Horn, Daniela Miao, and Michael Specter.
\newblock {Security Analysis of Wearable Fitness Devices (Fitbit)}.
\newblock \url{https://courses.csail.mit.edu/6.857/2014/files/17-cyrbritt-webbhorn-specter-dmiao-hacking-fitbit.pdf}, 2014.

\bibitem{Eric:2016}
Eric Clausing, Michael Schiefer, and Maik Morgenstern.
\newblock {AV TEST Analysis of Fitbit Vulnerabilities}.
\newblock Available at: \url{https://www.av-test.org/fileadmin/pdf/avtest_2016-04_fitbit_vulnerabilities.pdf}, 2016.

\bibitem{Schellevis}
Maarten Schellevis, Bart Jacobs, , and Carlo Meijer.
\newblock {Security/privacy of wearable fitness tracking IoT devices}.
\newblock Radboud University. Bachelor thesis: Getting access to your own Fitbit data., August 2016.

\bibitem{accenture}
{Accenture}.
\newblock Digital trust in the {IoT} era, 2015.

\bibitem{pwc}
{PwC 2016}.
\newblock Use of wearables in the workplace is halted by lack of trust.
\newblock  \url{http://www.pwc.co.uk/who-we-are/regional-sites/northern-ireland/press-releases/use-of-wearables-in-the-workplace-is-halted-by-lack-of-trust-pwc-research.html}.

\bibitem{Fereidooni:2017}
 Hossein Fereidooni, Tommaso Frassetto, Markus Miettinen, Ahmad-Reza Sadeghi, and Mauro Conti. 
\newblock {Fitness Trackers: Fit for Health but Unfit for Security and Privacy.}
\newblock {In Proc. IEEE International Workshop on Safe, Energy-Aware, \& Reliable Connected Health 
(CHASE workshop: SEARCH 2017), in press, Philadelphia, Pennsylvania, USA, July 17-19, 2017.}


\bibitem{Galileo}
{Galileo project}.
\newblock \url{https://bitbucket.org/benallard/galileo/}.

\bibitem{wireshark}
{Wireshark network protocol analyzer}.
\newblock \url{https://www.wireshark.org/}.

\bibitem{Andrew}
Andrew Hilts, Christopher Parsons, and Jerey Knockel.
\newblock {Every Step You Fake: A Comparative Analysis of Fitness Tracker Privacy and Security}.
\newblock Open Effect Report. \url{https://openeffect.ca/reports/Every_Step_You_Fake.pdf}, 2016.

\bibitem{Eric:2015}
Eric Clausing, Michael Schiefer, and Maik Morgenstern.
\newblock {Internet of Things: Security Evaluation of nine Fitness Trackers}.
\newblock AV TEST, The Independent IT-Security institue, Magdeburg, Germany, June 2015.

\bibitem{Zhou}
W.~Zhou and S.~Piramuthu.
\newblock {Security/privacy of wearable fitness tracking IoT devices}.
\newblock IEEE Iberian Conference on Information Systems and Technologies, 2014.

\bibitem{Rahman:2016}
Mahmudur Rahman, Bogdan Carbunar, and Umut Topkara.
\newblock {Secure Management of Low Power Fitness Trackers}.
\newblock Published in IEEE Transactions on Mobile Computing, Volume 15 Issue 2, Pages 447-459, February 2016.

\end{thebibliography}

\section*{Acknowledgments}
\vspace*{-0.5em}
Hossein Fereidooni is supported by the Deutsche Akademische Austauschdienst (DAAD). Mauro Conti is supported by the EU TagItSmart! Project (agreement H2020-ICT30-2015-688061) and IT-CNR/Taiwan-MOST 2016-17 ``Verifiable Data Structure Streaming". This work has been co-funded by the DFG as part of projects S1 and S2 within the CRC 1119 CROSSING, and by the BMBF within CRISP. Paul Patras has been partially supported by the Scottish Informatics and Computer Science Alliance (SICSA) through a PECE grant. 

We thank the Fitbit Security Team for their professional collaboration with us, and their availability to discuss our findings and address the vulnerabilities we identified.

\bibliographystyle{splncs}

\end{document}